\documentclass[twocolumn,preprintnumbers,amsmath,amssymb]{revtex4}
\usepackage{graphicx}
\usepackage{dcolumn}
\usepackage{bm}

\begin{document}
\title{How the surrounding water changes the electronic and magnetic properties of DNA}
\author{Julia Berashevich and Tapash Chakraborty}
\affiliation{Department of Physics and Astronomy, The University
of Manitoba, Winnipeg, Canada, R3T 2N2}
\begin{abstract}
Strong influence of water molecules on the transport and magnetic properties of 
DNA, observed in this study, opens up real opportunities for application of DNA
in molecular electronics. Interaction of the nucleobases with water molecules leads 
to breaking of some of the $\pi$ bonds and appearance of unbound $\pi$ electrons. 
These unbound electrons contribute significantly to the charge transfer at room 
temperature by up to 10$^3$ times, but at low temperature the efficiency of 
charge transfer is determined by the spin interaction of the two unbound electrons 
located on the intrastrand nucleobases. The charge exchange between the nucleobases 
is allowed only when the spins of unbound electrons are antiparallel. Therefore, 
the conductance of the DNA molecule can be controlled by a magnetic field. 
That effect has potential applications for developing a nanoscale spintronic 
device based on the DNA molecule, where efficiency of spin interaction will be 
determined by the DNA sequence. 
\end{abstract}
\maketitle

\section{Introduction}
The growing interest on the application of organic molecules for building nanoelectronic 
devices is motivated by several factors: conductance properties, self-assembly and 
molecular recognition properties. Self-assembly of molecular building blocks into 
well-structured systems allows us to exclude physical manipulation during fabrication 
of nanosize devices. Moreover, the structural damages of the electrical elements 
can be repaired without any physical contact using the molecular recognition. Since 
the DNA molecules satisfy all these requirements and, moreover, have been found to
conduct charge, its application to developing a nanoscale molecular devices 
is quite promising. While pristine DNA is not a good conductor, its conductivity
can be greatly enhanced by suitably changing the environment, thereby rendering
it an important element for the nanobio electronics \cite{revmod}.

An important finding that the overlapping of $\pi$ orbitals of the stacked base 
pairs can create a $\pi$-pathway for charge migration in DNA, has over the years 
inspired many research groups to investigate the electronic properties of this 
important biomolecule \cite{endres,ratner,lewis122,conw,gieN,bixon2,%
siguyam,voit,troisi,lewis,barton}. Two independent possible directions can be 
discerned from various experiments, reported as yet: the charge migration in DNA
\cite{lewis122,lewis,gieN,barton,lewisang,book} and its conductivity 
\cite{nanobio,yoo,kasumov,kawai,porath}. It is now more or less established that charge 
migration in DNA occurs through superexchange tunneling and charge hopping
\cite{ratner,jortner,berlin}. However, the experimental data related to DNA 
conductivity remains unclear: DNA molecules exhibit a wide range of behaviors, from 
insulator to metallic \cite{nanobio,yoo,kasumov,kawai,porath}. Only recently, this
issue has received a new twist, i.e., the humidity is recognized as an important 
factor controlling both DNA conductivity \cite{kawai1,other,tuuk} and DNA magnetic 
properties \cite{bouch,bouch1}. It has been observed that the DNA conductivity 
can increase exponentially by up to 10$^6$ times with rising humidity 
\cite{kawai1,other,tuuk}. In Ref.~\cite{tuuk}, participation of the DNA molecule 
in charge transport was verified by the high resistance of the environment, which 
exceeds up to 100 times the resistance of DNA itself. The origin of DNA conductivity
and its enhancement with humidity was not yet clearly understood. The current
interpretation of this phenomena \cite{other} rests on the change of DNA 
permittivity and therefore, the DNA conductivity due to adsorption of the water 
molecules on the DNA skeleton.

For DNA conductivity, the electronic interaction of the nearest-neighbor base 
pairs is the most important issue, that was extensively investigated by many 
research groups \cite{siguyam,voit,troisi,berash2,siebelas}. The parameters 
thus obtained were subsequently used for simulations of the charge transfer in 
DNA within different approaches, such as the tight-binding Hamiltonian 
\cite{book,macia,wang}, the system of kinetic equations \cite{jortner,berlin} 
and the polaron model \cite{conw,berash}. Interestingly, when the superexchange
tunneling and hopping were taken as two main transfer mechanisms, the experimentally 
observed features for hole migration \cite{gieN,lewisang} were indeed reproduced 
by the theories. However, these models fail to explain the diverse behaviors of DNA
conductivity observed in the experiments.

From our point of view, the inconsistency of theory and experiment on DNA 
conductivity lies with the reported theoretical approaches, where the charge 
transfer parameters were evaluated via the quantum-chemistry methods {\it in vacuum}
\cite{siguyam,voit,troisi,berash2,siebelas}. Moreover, most often this evaluation 
was performed for a single nucleobase in vacuum \cite{voit,troisi,siebelas} that 
disregards the significant shift of the highest occupied molecular orbital (HOMO)
energies due to the interstrand interaction of the nucleobases participating in the 
pair formation \cite{berash2,siguyam}. The discrepancy between the computed parameters 
and the experimental estimate is rather large. For example, the potential barrier for
the charge transfer from G-C to A-T has been estimated theoretically to be $\sim 0.7$ 
eV \cite{siguyam,berash2}, while the experimental value is 0.2 eV \cite{lewis}. 
Application of the pure electrostatic model to account for the solvation effect
\cite{berash1,new_my} has shown a decrease of this potential barrier from 0.7 eV to 
$\sim$ 0.5 eV, still higher than the experimental value. Therefore, it was suspected 
\cite{new_my} that interaction of DNA with water contributes not only to the
solvation effect but also by changing the nucleobase electronic properties due to 
their interactions with water.

Of late several papers have reported quantum-chemical calculations of the electronic 
properties of DNA surrounded by solvent molecules \cite{shuster,parinel,new_my}. We
have shown earlier \cite{new_my} that the nucleobases link to water and interactions 
between them can change the symmetry of the occupied $\pi$ orbitals localized on the 
nucleobases, that significantly shifts their orbital energies. Therefore, the
potential barrier for hole migration from G-C to A-T pairs is decreased from 0.7 eV 
(dehydrated pairs) to 0.123 (hydrated pairs). Moreover, occurrence of unbound $\pi$ 
electrons was found to be a reason for an increase of DNA conductivity with rising
humidity \cite{new_my}. The unbound $\pi$ electrons result in breaking of $\pi$ bonds, 
that occurs because of redistribution of the electron density from the $\pi$ bonds
toward the nitrogen atoms due to hydrogen bonding of nitrogen with water molecules. 
The increase of the charge transfer between two base pairs due to contribution of 
such unbound electrons (up to 250 times \cite{new_my}) thus found, is still much 
less than in the experiments \cite{kawai1,other,tuuk}. We suggest that the reason 
for the underestimation of the humidity effect lies in the fact that the DNA geometry 
transformation from B to A forms with dehydration of DNA is not taken into account.

In our present work we report on our investigation of the impact of humidity on
the orbital interaction and charge transfer between two base pairs if they are stacked 
according to the structural parameters of B-DNA and A-DNA. The aspects of the orbital 
interaction of the unbound electrons are considered also for an explanation of the 
magnetic properties of the B-DNA molecule, exhibiting the paramagnetic
behavior in a magnetic field \cite{bouch,bouch1}.

\section{Methods}
\begin{figure}
\includegraphics[scale=1.1]{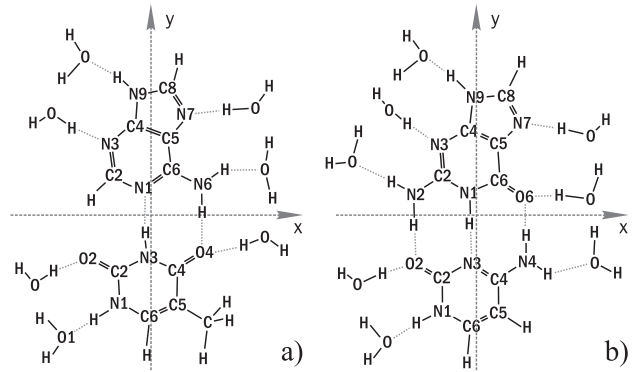}
\caption{The crystal structures: (a) water-(A-T) and
(b) water-(G-C) geometries. The water molecules are attached
by hydrogen bonds to the base pairs at positions given in
Ref.~\cite{gunst}.}
\label{fig:fig1}
\end{figure}

The molecular dynamics simulations for the DNA molecule placed into the ``water box" 
have shown that the nucleobases are able to make hydrogen bonding with the water 
molecules \cite{gunst}. For our investigation of the electronic properties of canonical
base pairs for dehydration and hydration, the optimized structures of the A-T and G-C 
base pairs without water molecules and the same structures connected by hydrogen bonds 
to the water molecules were constructed via the quantum-chemical methods. In the first 
stage, the geometries of the A-T and G-C base pairs were optimized in vacuum with the 
Jaguar program \cite{jaguar}. The Becke3-Lee-Yang-Parr functionals \cite{becke} and the 
restricted basis set with polarization and diffuse functions 6-31++G** were applied. 
Next, the crystal structure of the hydrated DNA base pairs were obtained by placing 
the water molecules close to the nucleobase atoms having the ability to make hydrogen 
bonding with the water molecules \cite{gunst}. The position of water molecules were 
optimized under the condition of frozen base pair geometries to save their planarity. 
The crystal structures of the hydrated (A-T) and (G-C) pairs thus obtained are shown 
in Fig.~\ref{fig:fig1}.

To build the poly-sequences such as (A-T)$_2$ and (G-C)$_2$ and also the mixed sequences, 
the optimized geometries of the hydrated (A-T) and (G-C) base pairs were stacked using 
the parameters of B-DNA and A-DNA. The natural bond orbital (NBO) analysis \cite{nbo} for 
these structures was performed, based on the electron density calculated with the 
B3LYP/6-31++G** functional, which was earlier found to be a better choice for this
purpose \cite{sun,new_my}. The application of the HF method for determination of the 
electron density distribution provides similar results as the NBO analysis \cite{new_my} 
except for weaker interaction of the natural bond orbitals. The $\Omega_i$ eigenfunctions 
are the natural bond orbitals built within the NBO analysis based on the input atomic 
orbital basis set obtained from the DFT. The advantage of the NBO analysis is that the 
Pauli exclusion principle is applied not only at the inner nodes that preserve orthogonality 
of the two electrons in the same orbital, but also at the outer nodes preserving the 
interatomic orthogonality. Within the NBO analysis \cite{nbo} the charge transfer between 
the nucleobases $Q_{DA}$ has been found as the sum of the $\Omega_i\rightarrow\Omega^*_j$ 
charge transfer between the donor orbital $\Omega_i$ belonging to one nucleobase to the 
acceptor orbital $\Omega^*_j$ on another nucleobase as
\begin{equation}
Q_{DA}=\sum_{i,j}Q_{i,j}=\sum_{i,j}q_i F_{i,j}^2/(\epsilon_i-\epsilon_j)^2
\label{eq:one}
\end{equation}
where $q_i$ is the donor orbital occupancy, $\epsilon_i,\epsilon_j$ are the orbital energies, 
$F_{i,j}$ is the off-diagonal element. The charge occupancy transfer is considered for 
stabilizing the orbital interaction, i.e. when the second order interaction energy $\Delta
E_{i,j}=-2F_{i,j}^2/(\epsilon_i-\epsilon_j)$ is characterized by the positive sign. The 
electronic coupling between the nucleobases has been estimated for the natural bond orbitals 
set as
\begin{equation}
V_{DA}=\sum_{i,j}V_{i,j}=\sum_{i,j}F_{i,j}^2/(\epsilon_i-\epsilon_j)
=\sum_{i,j}Q_{i,j}(\epsilon_i-\epsilon_j)/q_i.
\label{eq:two}
\end{equation}
To estimate the $\pi-\pi^*$ charge transfer in the DNA molecule, we took into account the 
charge exchange only between the $\pi$ orbitals.

\section{$\pi$-orbital interaction}
It is well known that charge transfer in DNA occurs due to the overlapping of the $\pi$ 
orbitals of the nearest-neighbor nucleobases. The efficiency of charge transfer is determined 
by the coupling of these $\pi$ orbitals and the difference of their energies. Thus, for two 
interacting base pairs, the transfer of a $\pi$ electron between stacked nucleobases is usually 
attributed to the interaction of HOMO and HOMO-1 orbitals (located on purines) and HOMO-2 and 
HOMO-3 (located on pyrimidines). The more efficient charge transfer is expected for two molecular 
orbitals located on the purines.

\begin{figure}
\makebox{a) }{\includegraphics[width=0.35\textwidth]{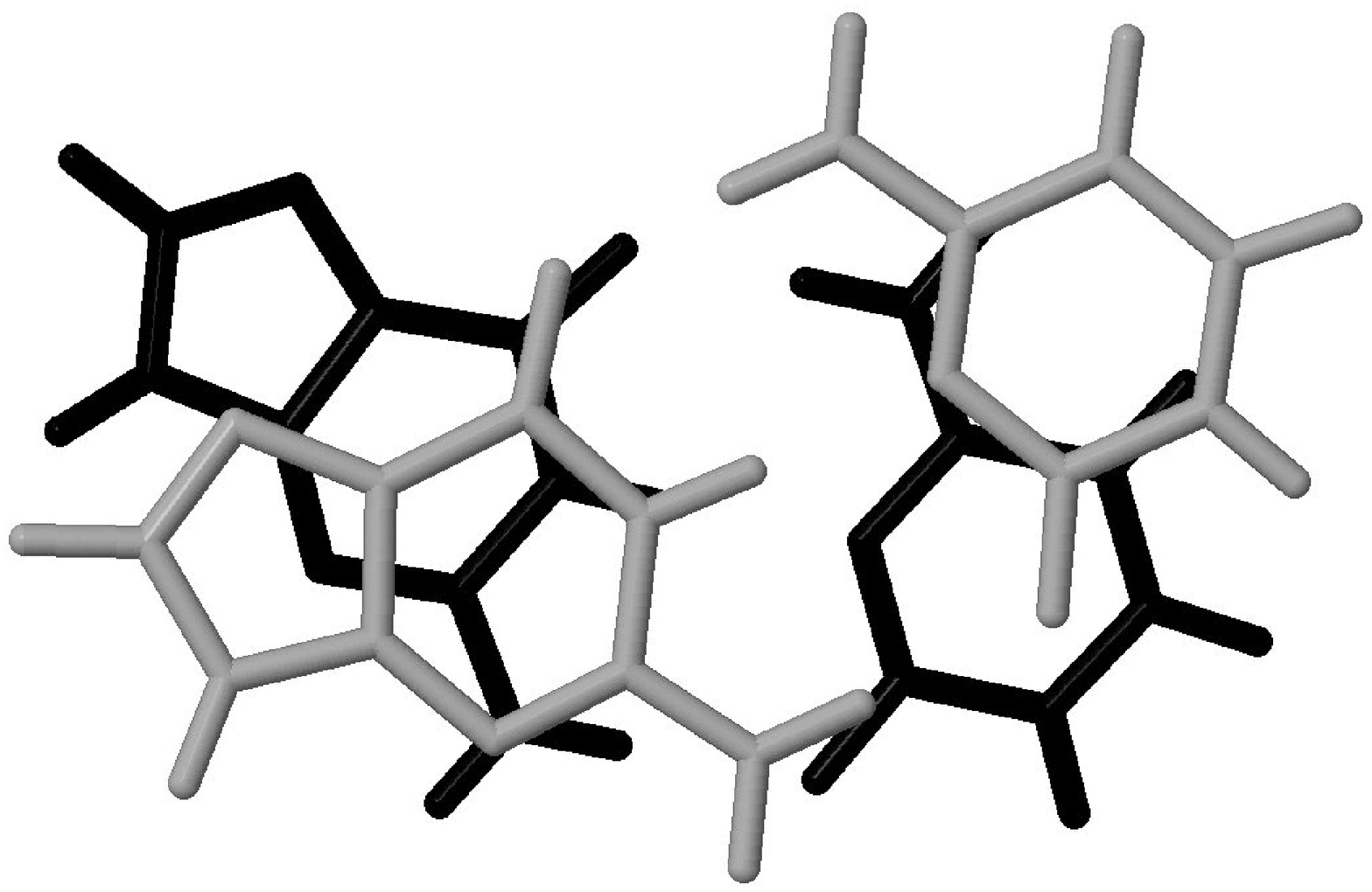}}
\makebox{b) }{\includegraphics[width=0.35\textwidth]{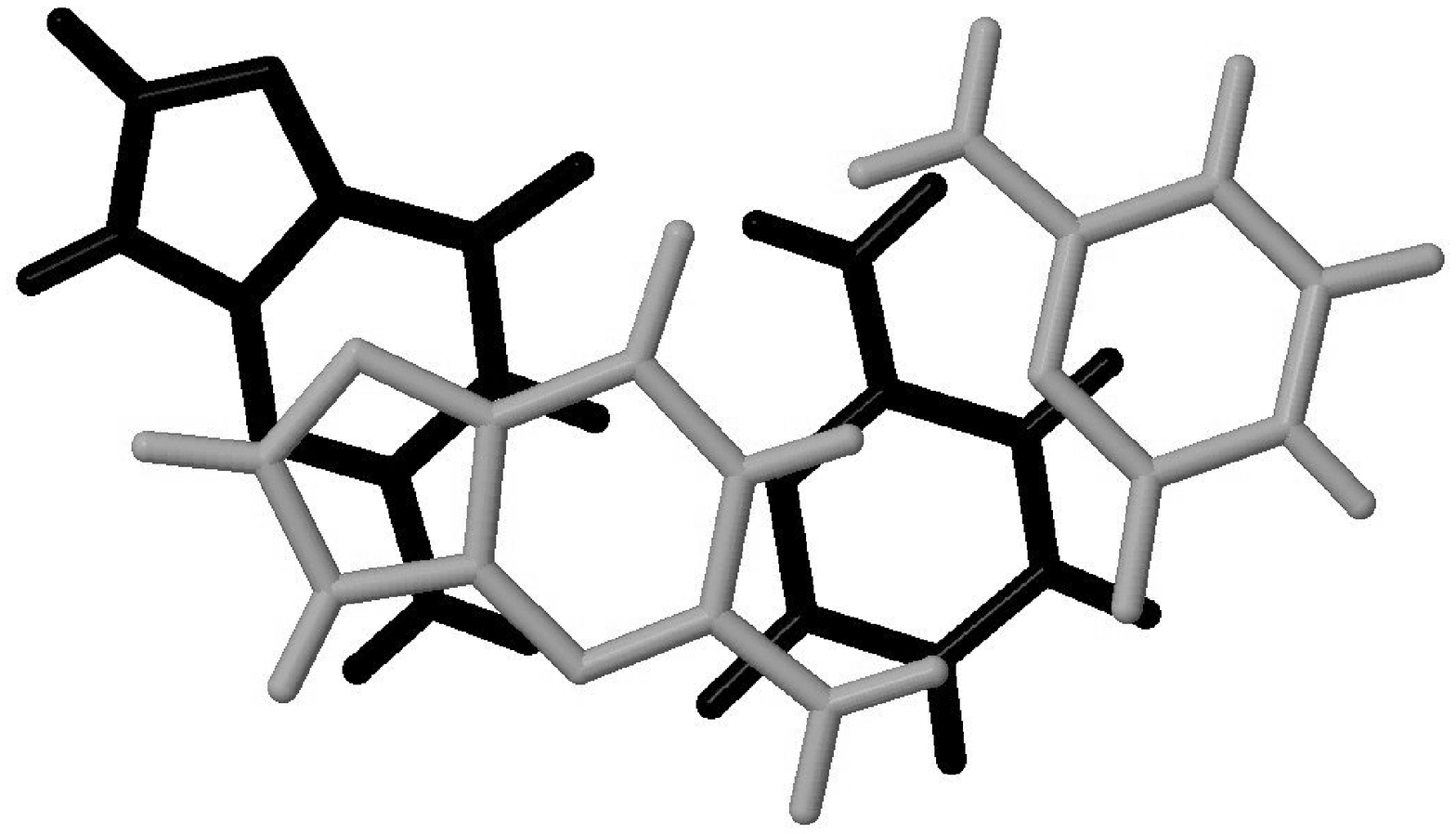}}
\caption{The geometry of the (G-C)$_2$ sequences where
the base pairs stacked with parameters characterized
a) for the B-DNA molecule and b) for the A-DNA molecule.}
\label{fig:fig2}
\end{figure}

In this section, we analyze the interaction of the HOMO and HOMO-1 orbitals in the system of 
two stacked base pairs. The efficiency of such interactions directly depends on the location 
of the base pairs relating to each other. It is known that sufficiently wet DNA is characterized 
by a B-form, while dehydration changes the geometry of the DNA molecule to the A-form. The 
distinction of A- and B-forms lies mostly in the helical twist $\theta$, which for
A-DNA is 32$^\circ$  and for B-DNA is 36$^\circ$, and in a dislocation of the base pairs from 
a helix axis $\Delta$, which for A-DNA is 4.5 \AA  and for B-DNA is -1.0 \AA \cite{book1}.
These parameters are important to describe a location of a base pair in the system of the 
two stacked pairs. The rest of the parameters, such as base pair tilt $\gamma$ and axial rise 
per nucleotide $d$, which are linearly dependent on each other \cite{book1}, do not significantly 
change this location. The geometry of the (G-C)$_2$ sequences, where base pairs are stacked
with structural parameters of B-DNA and A-DNA, are presented in Fig.~\ref{fig:fig2}. As the 
nucleobases in the B-form are located on top of each other (the displacement from the helix 
axis $\Delta$ is close to zero), the most important contribution to the base pair interaction 
comes from the intrastrand part. For the A-form, the dislocation of the base pair from the helix 
axis causes both the intra- and interstrand charge exchange between the nucleobases.

Here we study the influence of the base pair location in a system of two stacked base pairs, 
such as (A-T)$_2$ and (G-C)$_2$ sequences. The shift of the $\pi$-orbital energies due to the
interaction between two stacked base pairs can be determined from
\cite{rauk}
\begin{equation}
e_{\psi_1}\approx e_{\psi^0_1}-H_{12}+(e_{\psi^{0}_1}-H_{12})S_{12}
\label{eq:three}
\end{equation}
\begin{equation}
e_{\psi_2}\approx e_{\psi^0_2}+H_{12}-(e_{\psi^0_2}+H_{12})S_{12}
\label{eq:four}
\end{equation}
where $e_{\psi^{0}_{1,2}}$ and $e_{\psi_{1,2}}$ are the $\pi$-orbital energies respectively 
before interaction and after interaction, $H_{12}=\langle\psi_1\mid h^{eff}\mid\psi_2\rangle$ 
is the intrinsic interaction integral and $h^{eff}$ is the effective core potential.

\begin{figure}
\makebox{\includegraphics[width=0.48\textwidth]{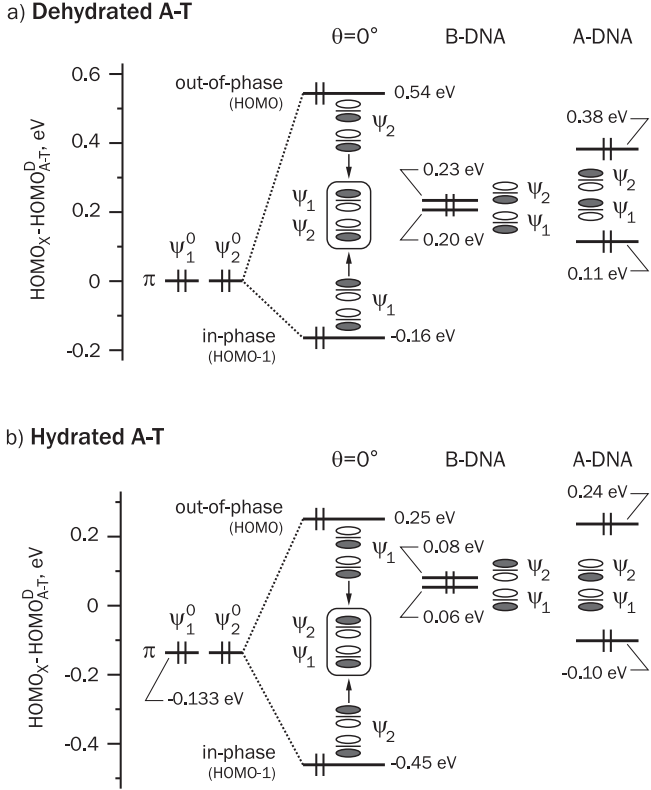}}
\caption{The shift of the HOMO and HOMO-1 energies and change of
the $\psi_{1}$ and $\psi_{2}$ wavefunctions' symmetry due to the
stacking of the two dehydrated A-T pairs (a)
($e_{\psi^{0}_{1(2)}}$) or two hydrated A-T pairs (b) into
(A-T)$_2$ sequences with different parameters: 1) two parallel
base pairs ($\gamma$=0$^\circ$, $d$=3.4 \AA) stacked with twist
angle of $\theta$=0$^\circ$; 2) two base pairs staked with
parameters corresponding to B-DNA ($\gamma$=0$^\circ$, $d$=3.4
\AA, $\theta$=36$^\circ$) and 3) with parameters corresponding
to A-DNA ($\gamma$=10$^\circ$, $d\approx$ 3.0 \AA,
$\theta$=32$^\circ$). All energies for the (A-T)$_2$ sequences are
calculated with respect to HOMO energy of the single dehydrated
A-T pair ($e_{\psi^{0}_{1(2)}}$).} \label{fig:fig3}
\end{figure}

\begin{figure}
\makebox{\includegraphics[width=0.48\textwidth]{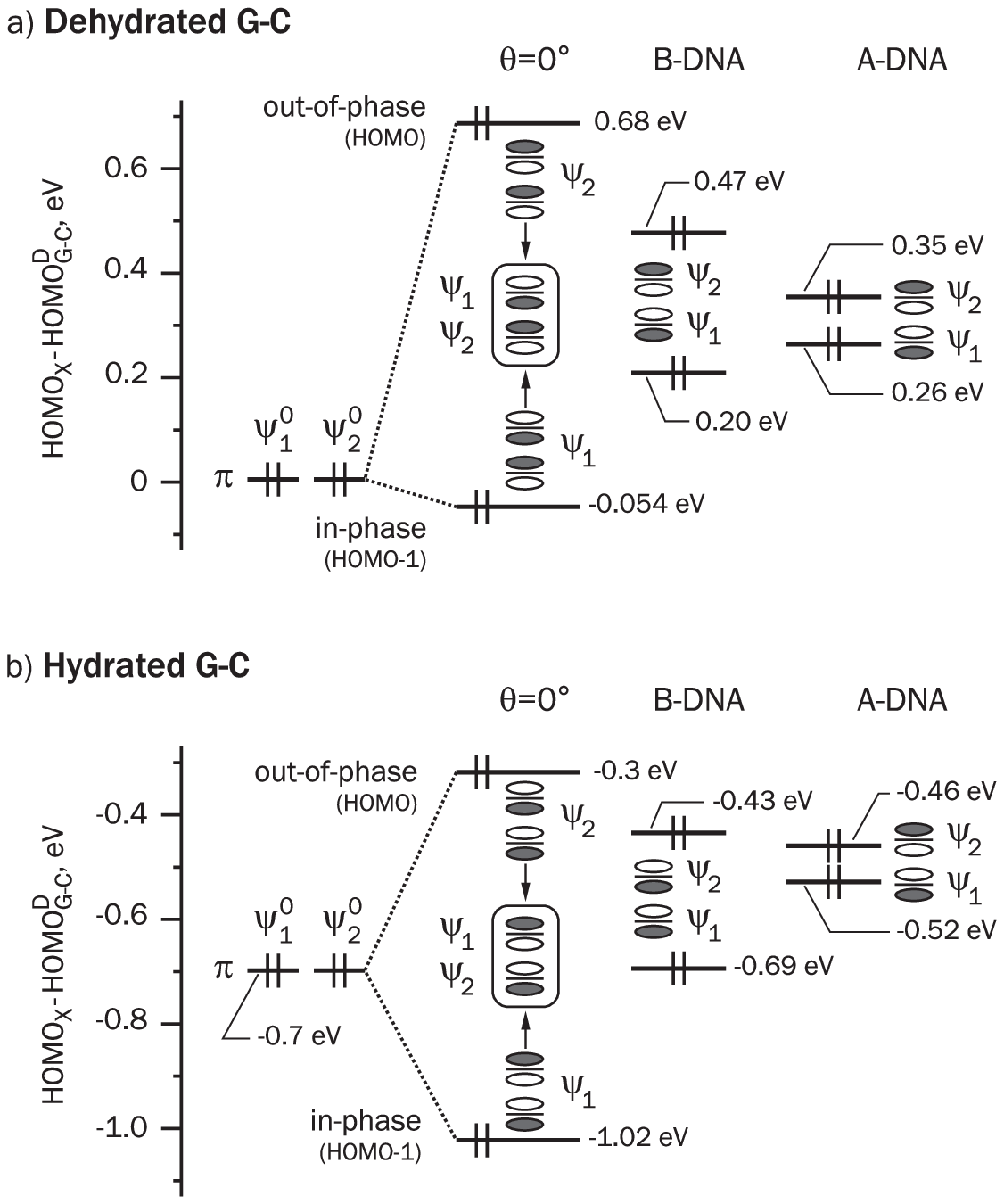}}
\caption{The shift of the HOMO and HOMO-1 energies and change of
the $\psi_{1}$ and $\psi_{2}$ wavefunction's symmetry due to the
stacking of the two dehydrated G-C pairs (a)
($e_{\psi^{0}_{1(2)}}$) or two  hydrated G-C pairs (b) into
(G-C)$_2$ sequences with different parameters (for parameters see
points 1-3 in caption for Fig.~\ref{fig:fig3}). All energies for
the (G-C)$_2$ sequences are calculated with respect to the HOMO
energy of the single dehydrated G-C pair ($e_{\psi^{0}_{1(2)}}$).}
\label{fig:fig4}
\end{figure}

In the system of (A-T)$_2$ and (G-C)$_2$ sequences the interaction of filled $\pi$ orbitals 
is repulsive and therefore, the base pairs tend to repeal each other \cite{rauk}. This process 
is dominant when filled orbitals are orthogonalized, i.e. when the tilt angle $\gamma$=0$^\circ$ 
and the twist angle $\theta$=0$^\circ$. In this case, because the wavefunctions $\psi_1$ 
and $\psi_2$ are interacting in such a way that their lobes of the same sign overlap 
($\psi_1$ and $\psi_2$ are orthogonal orbitals, i.e. {\it in-phase} interaction), this leads 
to a decrease of the charge occupancy transfer $Q_{DA}$ between the intrastrand nucleobases 
and the Pauli repulsion becomes dominant. The result of such an interaction is a shift of the 
$\pi$-orbital energies $e_{\psi_{1}}$ and $e_{\psi_{2}}$ by $H_{12}$, while the contribution 
of the $(e_{\psi^{0}_1}\pm H_{12})S_{12}$ term is approximately zero. The shift of the HOMO 
and HOMO-1 orbital energies after perturbation is presented in Fig. ~\ref{fig:fig3} for
the (A-T)$_2$ sequence and in Fig.~\ref{fig:fig4} for the (G-C)$_2$ sequence. For a twist 
angle of 0$^\circ$ the $\psi_i$ wavefunctions are almost equally delocalized over two purines and
the splitting of the HOMO and HOMO-1 orbitals is large $\sim
2H_{12}$.

However, in the A- and B-DNA molecule the pair bases are stacked with a helical 
twist and additionally they are shifted from the helix axis. Therefore, the $\pi$ 
electron density of the $\psi_1$ and $\psi_2$ wavefunctions are shifted such that it
destroys the wavefunction orthogonality and results in a rise of the intrastrand 
charge occupancy transfer $Q_{DA}$ between the stacked base pairs. Depending on the 
efficiency of the charge occupancy transfer, which is determined by the orbital overlap
$S_{12}$ and symmetry of the interacting orbitals, the orbital energies $e_{\psi_1}$ and 
$e_{\psi_2}$ can be significantly shifted in comparison to the orthogonal case (see the 
case of $\theta$=0$^\circ$ in Figs.~\ref{fig:fig3}--~\ref{fig:fig4}). Since the $\pi$-electron 
clouds are shifted, the charge occupancy transfer is significant even when two orbitals
$\psi_1$ and $\psi_2$ are in-phase (see dehydrated B-DNA in Fig.~\ref{fig:fig4} (a)). 
Therefore, for the B-DNA and the A-DNA, the wavefunctions $\psi_1$ and $\psi_2$ are strongly 
localized on one nucleobase and splitting of their HOMO and HOMO-1 orbitals is decreased 
relating to orthogonal case due to the contribution from the $(e_{\psi^0_1}\pm H_{12})S_{12}$ 
term in Eqs.~\ref{eq:three}--~\ref{eq:four}. In our earlier works \cite{new_my}, the 
humidity was found to change the electron density distribution and symmetry of the $\pi$ 
orbitals for the single G-C and A-T base pairs. Here, we observe that for the (A-T)$_2$  
and (G-C)$_2$ sequence in the B-form, hydration reverses the interaction of the $\psi_1$ 
and $\psi_2$ wavefunctions. For example, for the (A-T)$_2$ sequence, the hydration changes 
interaction from out-of-phase to in-phase, while for the (G-C)$_2$ sequence from in-phase 
to out-of-phase. For the A-DNA the humidity changes the symmetry of the $\psi_1$ and
$\psi_2$ wavefunctions as well, but preserve the type of their interaction, i.e. the 
interaction of the $\psi_1$ and $\psi_2$ wavefunctions always remains out-of-phase for 
the (A-T)$_2$ sequence and in-phase for the (G-C)$_2$ sequence.

The symmetry of the $\psi_1$ and $\psi_2$ wavefunctions localized on purines shows 
an opposite behavior for the (A-T)$_2$ and (G-C)$_2$ sequences staked with parameters of
the B- and A-forms. This was found to be a result of the interaction of purines with 
pyrimidine within the DNA pairs. For two stacked dehydrated guanines (G)$_2$ and two 
stacked dehydrated adenines (A)$_2$ only out-of-phase interactions of their $\psi_1$ 
and $\psi_2$ wavefunctions corresponding to HOMO and HOMO-1 orbitals are observed for 
both A- and B-forms in dehydrated case. Therefore, the stacked guanines and the adenines 
originally are characterized by the same symmetry of the $\psi_1$ and $\psi_2$ wavefunctions. 
Their properties are changed due to the contribution of the pyrimidine bases participating 
in the base pair formation. Thus, for the (G-C)$_2$ sequences, the contribution of the $\pi$ 
orbital symmetry from cytosine switch the interaction of the $\psi_1$ and $\psi_2$ wavefunctions 
to in-phase type, while the contribution of the $\pi$ orbital symmetry from thymine for the 
(A-T)$_2$ sequence do not provide such a phase reversal. Similar to the occupied orbitals, 
the (A-T)$_2$ and (G-C)$_2$ sequences have shown the opposite type of interaction of the 
lowest unoccupied molecular orbitals (LUMOs), such as LUMO and LUMO-1.

We thus conclude that the transport properties of the poly(dG)-poly(dC) and poly(dA)-poly(dT) 
will be quite different due to the difference of their electronic structures, namely due
to different symmetry of their interacting HOMOs and LUMOs. This implies that experimentally 
observed \cite{yoo} different properties of poly(dG)-poly(dC) and poly(dA)-poly(dT) molecules 
showing respectively the $p$-type and $n$-type conductance can be related to their intrinsic 
electronic properties.

\section{$\pi$ charge transfer}
Our next step is the quantitative characterization of the charge occupancy transfer between 
stacked base pairs performed with the NBO analysis for DNA of B and A forms with different 
levels of humidity. For the hydrated (A-T)$_2$ and (G-C)$_2$ sequences in the B-form, the 
intrastrand interaction between the stacked pairs and interaction of the nucleobases with 
water molecules drastically change the electron density distribution and population of the 
$\pi$ orbitals over the base pairs. As a result, the covalent structure of the nucleobases 
is converted to a structure with separated charge (ionic), that has been obtained by the 
NBO analysis \cite{new_my}. In particular, for adenine and guanine the $\pi$ electron 
density between the $C4$ and $C5$ atoms is shifted. Therefore, instead of a double bonding 
to the $C5$ atom, the $C4$ atom {\it donates} an electron to form a double bond with the
$N9$ atom, which contributes one $\pi$ electron from the lone pair. Consequently, the $C5$ 
atom has only three bonds and carries a negative charge (one $\pi$ electron not locked up 
to the covalent bond), while the $N9$ atom has four bonds (one missing $\pi$ electron) 
and carries a positive charge. The charge separation occurs for the cytosine in a similar
manner, where the $N3$ atom carries a negative charge and the $N4$ atom carries a positive
charge. Interestingly, for the (A-T)$_2$ sequence, the adenines are converted to the 
ionic structure already for the dehydrated case, while thymine always has a covalent 
structure. The origin of this behavior is the low weighting of the covalent structure of
adenine and the high weighting of thymine \cite{sun}.

The alteration of the double bond pattern due to conversion of the covalent structure 
to the structure with separated charges changes the orbital perturbation and the intrastrand 
charge occupancy transfer between the nucleobases within each base pair. Therefore, the 
interaction of guanine and cytosine within the hydrated G-C pair ($\gamma$=0) is significantly 
increased in comparison to that for the dehydrated G-C pair. Their interstrand interactions 
are enhanced because of the change of the covalent structure of guanine and cytosine to the 
ionic structure and consequently to lowering the energy gap between their $\pi$ orbitals. In 
the covalent structure of the dehydrated G-C pair the charge transfer ($Q_{\rm{G}\rightarrow 
\rm{C}}$=0.00043 \={e}) occurs mostly between the $N1$ lone pair (guanine) and the $\pi^*$ 
orbital of the $N3$-$C4$ bond (cytosine), where the energy gap is $\epsilon_{N1(\rm{G})}
-\epsilon_{N3-C4(\rm{C})}= 5.98$ eV. In the structure with separated charge, the charge 
transfer occurs mostly between the lone pair of the $N1$ atom (guanine) and unbound $\pi$
electron on the $N3$ atom (cytosine). The energy gap between them is decreased to 
$\epsilon_{N1(\rm{G})}-\epsilon_{N3(\rm{C})}= 1.9$ eV and the amount of charge transfer 
is $Q_{\rm{G}\rightarrow\rm{C}}$=0.0054 \=e. The electronic coupling between guanine and
cytosine is estimated to be $V_{\rm{G}\rightarrow \rm{C}}$=0.0065 eV for the hydrated case 
against $V_{\rm{G}\rightarrow\rm{C}}$=0.0016 eV for the dehydrated. Unlike in dry DNA
in wet DNA the interstrand charge transfer within the G-C pair can actually make a large 
contribution to the DNA transport properties because of a stronger electronic coupling
between guanine and cytosine and a lower energy gap between them. The enhancement of the 
interaction of the thymine and adenine within a A-T pair due to hydration is found to be 
very small.

Within the ionic structure of both guanine and adenine, the unbound $\pi$ electrons are 
localized on the $C5$ atom. However, the participation of these $\pi$ electrons in the 
intrastrand charge transfer within the (G-C)$_2$ and (A-T)$_2$ sequences is completely 
different for these purines due to the symmetry of the interacting orbitals HOMO and 
HOMO-1 observed in the previous section (in-phase and out-of-phase). For the (G-C)$_2$ 
sequence in B form, the $\psi_1$ and $\psi_2$ wavefunctions are out-of-phase and therefore, 
the charge transfer between two guanines is large because of the permitted direct charge 
exchange between the $C5$ atoms ($Q_{C5(\rm{G_1})\rightarrow}$$_{C5(\rm{G_2})}$) with
participation of the unbound $\pi$ electrons. The overlap of these unbound electrons 
for the B-DNA is significant (the overlap matrix $S_{C5(\rm{G_1})\rightarrow}$
$_{C5(\rm{G_2})}$=0.0245) and energy gap for transfer is rather small $\epsilon_{C5(\rm{G_1})}
-\epsilon_{C5(\rm{G_2})}=0.33$ eV. All of these provide a charge occupancy transfer from 
one guanine to another with participation of the unbound electrons to be $Q_{C5(\rm{G_1})
\rightarrow}$$_{C5(\rm{G_2})}$=0.145 \={e}, that is a dominant charge transfer channel for 
the (G-C)$_2$ sequence. Because the unbound $\pi$ electrons on the $C5$ atoms are strongly
localized, their orbital overlapping is sensitive to the twist angle. Therefore, the
$Q_{C5(\rm{G_1})\rightarrow}$$_{C5(\rm{G_2})}$ charge transfer exist only in the confined 
range of twist angle from $\theta$=33$^\circ$ to $\theta$=44$^\circ$. Our computational
results for the charge occupancy transfer between guanines and cytosines within the (G-C)$_2$ 
sequence are presented in Table~\ref{tab:table1}. Dehydration of the (G-C)$_2$ sequence 
in B-form leads to a major decrease of the charge occupancy transfer and change of its 
trend as result of application of the symmetry rule (see dehydrated B-DNA in Fig.~\ref{fig:fig4}). 
For the intrastrand cytosines in the (G-C)$_2$ sequences, the dominance of the $N3$-$C4(\rm{C_1})$ 
$\rightarrow C4$-$N4(\rm{C_2})$ interaction in the dehydrated case is switched to the 
prevalence of the $N3$-$C4(\rm{C_1})\rightarrow$$N3$-$C4(\rm{C_2})$ interaction in
the hydrated case, that significantly increases the charge transfer between the hydrated 
bases. However, contribution of the cytosine in the charge occupancy transfer between stacked 
G-C pairs is quit small and can be neglected for both B and A-forms. For the (G-C)$_2$ 
sequence stacked with parameters corresponding to the A-DNA, their HOMO and HOMO-1 are 
in-phase that invalidates the charge transfer between natural bond orbitals created by
unbound electrons and decrease substantially the whole intrastrand charge transfer between 
guanines. As was already mentioned above, sufficiently wet DNA is characterized by the B-form 
and dry DNA is of A-form, and the poly(dG)-poly(dC) conductivity with humidity is 
enhanced 10$^3$-10$^6$ times \cite{kawai1,other,tuuk}. The comparison of the magnitude 
$Q=Q_{DA}-Q_{AD}$ computed for the hydrated B-DNA(wet DNA) (see Table ~\ref{tab:table1}) and
dehydrated A-DNA is calculated to be $\sim$ 10$^3$ that is in excellent agreement with 
the experimental data \cite{kawai1}. As was expected, the charge occupancy transfer observed 
between interstrand guanine and cytosine belonging to different base pairs is quite large 
($Q_{\rm{G_{1}}\rightarrow \rm{C_{2}}}$=0.0322\={e}) for the A-form in comparison to that for
the B-DNA structure ($Q_{\rm{G_{1}}\rightarrow \rm{C_{2}}}$=0.00136 \={e}).

\begin{table}
\caption{\label{tab:table1} The intrastrand charge transfer between guanines $\rm{G_{1}}
\rightarrow \rm{G_{2}}$ and cytosines $\rm{G_{1}}\rightarrow \rm{G_{2}}$ within (G-C)$_2$ 
sequences under different conditions.}
\begin{ruledtabular}
\begin{tabular}{c|c|c|c}
 & $Q_{DA}$, \={e} & $Q_{AD}$, \={e} & $Q=Q_{DA}-Q_{AD}$, \={e} \\
\hline
\multicolumn{4}{c}{$\rm{G_{1}}\rightarrow \rm{G_{2}}$} \\
\hline
hydrated  B-DNA & 0.1630 & 0.0063  & 0.1567  \\
dehydrated  B-DNA  & 0.0030 & 0.0036 & -0.0007 \\
hydrated  A-DNA   & 0.0062 & 0.0140 & -0.0078 \\
dehydrated  A-DNA  & 0.0017 & 0.0015 & 0.00017 \\
\hline
\multicolumn{4}{c}{$\rm{C_{1}}\rightarrow \rm{C_{2}}$} \\
\hline
hydrated  B-DNA & 0.0034 & 0.0030  & 0.0003 \\
dehydrated  B-DNA  & 0.0003 & 0.0010 & -0.0007 \\
hydrated  A-DNA   & 0.0100 & 0.0069  & 0.0033 \\
dehydrated  A-DNA  & 0.0042 & 0.0046 & -0.0004 \\
\end{tabular}
\end{ruledtabular}
\end{table}

For adenines within the (A-T)$_2$ sequences stacked according to parameters of the B-form, 
the overlap of the unbound $\pi$ electrons is $S_{C5(\rm{A_1})\rightarrow}$$_{C5(\rm{A_2})}
=-0.0265$, that is similar to that for the guanines, but $\psi_1$ and $\psi_2$ wavefunctions 
are in-phase which restricts the $C5(\rm{A_1})\rightarrow$$C5(\rm{A_2})$ charge transfer. 
However, the unbound $\pi$ electrons participate in the intrastrand charge exchange with 
$\pi^*$ orbitals on the $C4$-$N9(\rm{A_2})$ bond, but the efficiency is not high,
$Q_{C5(\rm{A_1})\rightarrow}$$_{C4-N9(\rm{A_2})}$=0.0223 \={e}, because of a large energy 
gap $\epsilon_{C5(\rm{A_1})}-\epsilon_{C4-N9(\rm{A_2})}= 1.9$ eV. Our computational results 
for the (A-T)$_2$ sequences are presented in Table~\ref{tab:table2}. For adenines within the
hydrated (A-T)$_2$ sequences of A-form, the HOMO orbitals are out-of-phase and $C5(\rm{A_1})
\rightarrow C5(\rm{A_2})$ charge transfer is permitted. However, the overlap matrix of this 
transfer is small, $S_{C5(\rm{A_1})\rightarrow}$$_{C5(\rm{A_2})}=0.0130$ and the magnitude 
of the off-diagonal element $F_{C5(\rm{A_1})\rightarrow}$$_{C5(\rm{A_2})}$ is 10 times 
smaller than that for B-DNA. Therefore, the contribution of the charge exchange between 
unbound electrons in (A-T)$_2$ sequences of the A-form is less. Because thymine structure 
remains covalent independent of hydration, the charge transfer between thymines for the 
A and B-form is slightly modified by hydration only due to a change of the $(\epsilon_i-
\epsilon_j)$ energy gap. Finally, an increase of the $\pi-\pi^*$ charge transfer for the 
(A-T)$_2$ sequences is found to be only $\sim$ 40 times due to the hydration, while 
the experimental value was 10$^3$ times \cite{kawai1}. This discrepancy is related to the 
applicability of the simulated results only for low temperature range, where orbital 
symmetry of the unbound electrons is preserved, i.e. when the energy required to change 
the orbital symmetry is $\Delta E_S$ $\ll k_{B}T/q$. We estimated the magnitude of 
$\Delta E_S$ for adenines within the (A-T)$_2$ sequence as the difference between the 
$\epsilon_{C5(\rm{A_1})}-\epsilon_{C5(\rm{A_2})}$ for unbound electrons when the 
wavefunctions are out-of-phase and in-phase, which was found to be $\Delta E_S$=0.12 eV. 
Therefore, already at a low temperature ($T <$100 K) unbound $\pi$ interacting electrons
have enough energy to switch the wavefunctions from out-of-phase to in-phase, which
opens up the opportunity for a direct $C5(\rm{A_1})\rightarrow$$C5(\rm{A_2})$ charge transfer 
in the (A-T)$_2$ sequences of the B-form. As a result, at room temperature the charge 
transfer with participation of the unbound electrons in B-DNA is possible not only for 
the (G-C)$_2$ sequence but also for the (A-T)$_2$ sequence. Therefore, at room
temperature the hydration of DNA should provide an increase of the DNA conductance 
approximately as the same magnitude for the poly(dG)-poly(dC) and poly(dA)-poly(dT) chains 
(up to 10$^3$ times), because the orbital overlap of nearest-neighbor unbound electrons for 
stacked guanines and stacked adenines is the same ($S_{C5(\rm{G_1(A_1}))\rightarrow}$
$_{C5(\rm{G_2(A_2}))}\approx\pm$0.025) and orbital symmetry is not preserved at room 
temperature.

\begin{table}
\caption{\label{tab:table2} The intrastrand charge transfer between adenines $\rm{A_1}
\rightarrow \rm{A_2}$ and thymines $\rm{T_1}\rightarrow \rm{T_2}$ within (A-T)$_2$ 
sequences under different conditions.}
\begin{ruledtabular}
\begin{tabular}{c|c|c|c}
 & $Q_{DA}$, \={e} & $Q_{AD}$, \={e} & $Q=Q_{DA}-Q_{AD}$, \={e} \\
\hline
\multicolumn{4}{c}{$\rm{A_1}\rightarrow \rm{A_2}$} \\
\hline
hydrated  B-DNA & 0.0269 & 0.0052 & 0.0218 \\
dehydrated  B-DNA  & 0.0040 & 0.0026 & 0.0014 \\
hydrated  A-DNA   & 0.0053  & 0.0140 & -0.0087  \\
dehydrated  A-DNA  & 0.0025 & 0.0055 & -0.0030\\
\hline
\multicolumn{4}{c}{$\rm{T_1}\rightarrow \rm{T_2}$} \\
\hline
hydrated  B-DNA & 0.0020 & 0.0010 & 0.0010\\
dehydrated  B-DNA  & 0.0018 & 0.0011  & 0.0007 \\
hydrated  A-DNA   & 0.0056 & 0.0033 & 0.0023 \\
dehydrated  A-DNA  & 0.0057 &0.0033 & 0.0024 \\
\end{tabular}
\end{ruledtabular}
\end{table}

In summary, we found that the occurrence of the unbound $\pi$ electrons activated 
by humidity suppresses the DNA band gap from $\sim$ 8.0 eV for the dehydrated DNA 
to $\sim$ 3.0 eV for the hydrated DNA. This value of the band gap was often observed 
in the experiments \cite{fiebig,frohne}. Therefore, the occurrence of the unbound 
$\pi$ electrons suppressing the band gap of wet DNA is the main factor for high B-DNA 
conductivity.

\section{DNA magnetism}
A noninvasive means to investigate the intrinsic electronic properties of DNA was
recently applied to wet and dry $\lambda$-DNA by measuring the magnetization of 
this molecule at different humidity levels \cite{bouch}. The paramagnetic behavior
of $\lambda$-DNA in the B-form was observed at low temperature, while dry A-form was 
shown to be diamagnetic for a wide temperature range. The solvent of distilled $\rm{H_2O}$ 
itself is paramagnetic due to the contribution from the O$_2$ molecules, but the measured 
magnitude of magnetization of distilled water was not large enough in comparison to 
that for wet $\lambda$-DNA \cite{bouch1}, which proves the origin of the DNA paramagnetism
from the intrinsic DNA properties. The orbital motions of free electrons have been 
attributed to be the reason of this paramagnetic effect, while the spin contribution to 
paramagnetism was claimed to be unlikely. However, the interaction of nearest-neighbor 
unbound $\pi$ electrons created by interaction of the base pairs with water molecules 
can be a source of spin paramagnetism, because of possible pairing of the unbound
electrons with parallel or anti-parallel spins.

\begin{figure}
\makebox{\includegraphics[width=0.40\textwidth]{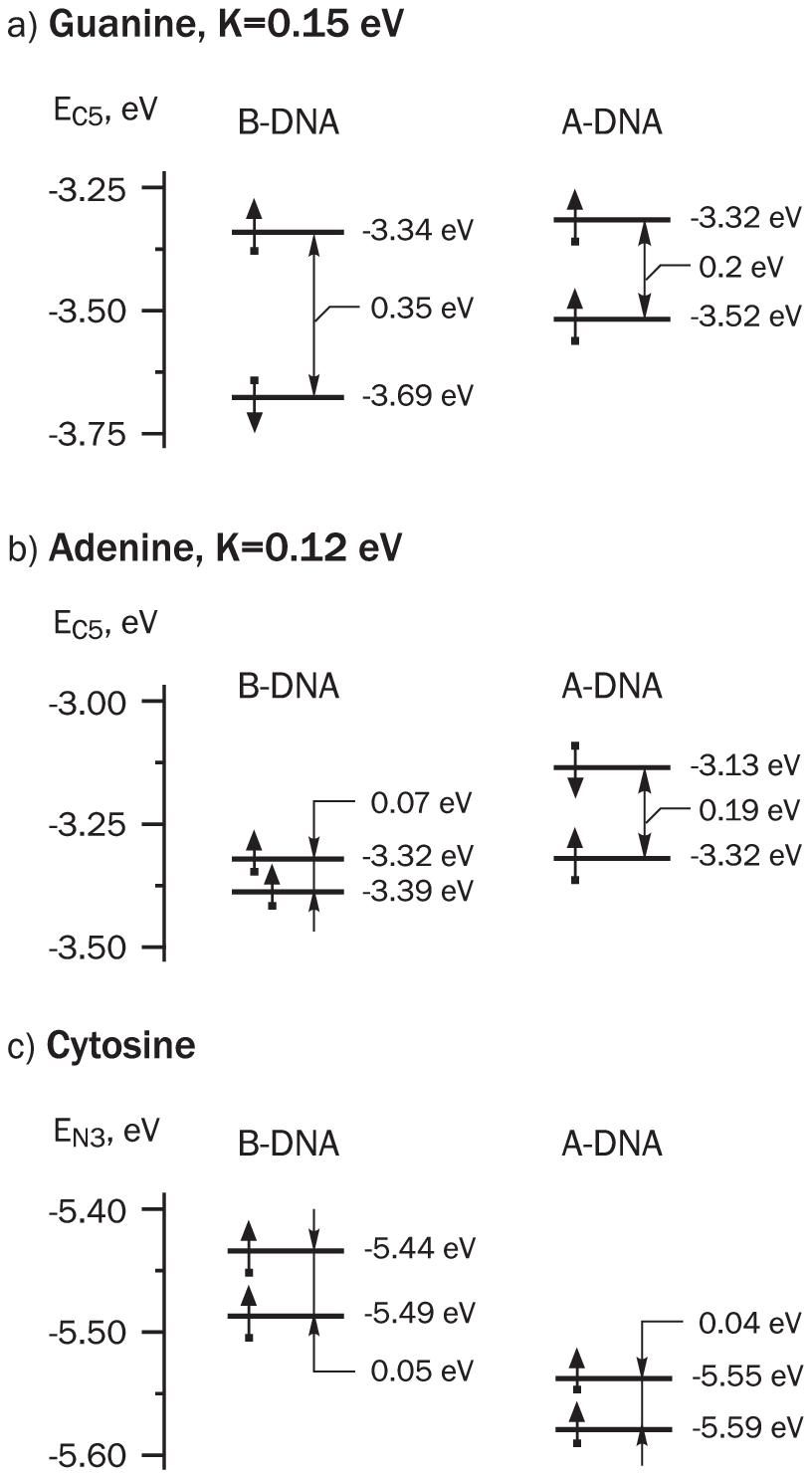}}
\caption{The energetics of the natural bond orbitals $\Omega_i$
corresponding to the valence unbound electrons formed within the
nucleobases due to their interaction with water a) for guanines
within (G-C)$_2$ sequence, b) for adenines within (A-T)$_2$
sequence and c) for cytosines within (G-C)$_2$ sequence. $K$ is
the exchange repulsion term calculated as difference of energy
$\Delta E$ for paired electrons with anti-parallel spins and
parallel spins.} \label{fig:fig5}
\end{figure}

We have shown in the previous section that each DNA base pairs linked to water molecules 
has unbound $\pi$ electrons, whose interaction is not strong because of the large separation 
(3.4 \AA). The consideration of the Pauli exclusion principle for outer nodes (interatomic 
orthogonality) in the NBO analysis allows us to analyze the spin state of unbound electrons. 
Originally, the phase of the interacting wavefunctions is taken from the standard
calculation B3LYP/6-31++G** within the restricted basis set. If the wavefunctions of the 
HOMO and HOMO-1 orbitals obtained within the DFT method are in-phase, then the unbound electrons 
obtained within the NBO analysis are also in-phase (see Figures~\ref{fig:fig3} and ~\ref{fig:fig4}), 
i.e., paired with parallel spin. Their natural orbitals are closer in energy because of
absence of exchange repulsion term $K$ in the interaction energy of two electrons. For the 
two unbound electrons paired with anti-parallel spins (out-of-phase interaction of HOMO
wavefunctions) their orbital energy splitting is quite large. The energy diagram of the 
unbound electrons in the guanines within the (G-C)$_2$ sequence, in adenines within the 
(A-T)$_2$ sequence and in cytosines within the (G-C)$_2$ sequence are presented in Fig.
\ref{fig:fig5} a), b) and c) respectively for hydrated sequences in B and A-forms. Two 
stacked cytosines are found to be source of paramagnetic behavior because their paired 
unbound electrons have parallel spin, while for the stacked adenines and guanines the
interaction of the unbound electrons largely depend on the DNA geometry. However, the 
stacked adenines as are also expected to contribute to the B-DNA paramagnetism.

Therefore, with regards to the diamagnetic and paramagnetic properties of DNA, dry DNA 
is expected to be always diamagnetic because of the absence of unbound electrons, while 
wet DNA should be paramagnetic and the efficiency of paramagnetism is directly dependent 
on the DNA sequences. The control of the spin of the unbound electrons by the magnetic 
field would open up real opportunities for using DNA in spintronics \cite{spintro}.

\section{Conclusion}
Our findings that hydration of DNA activates the occurrence of the unbound $\pi$ electrons, 
which can easily participate in conductance because of narrowing of the band gap to $\sim$ 
3.0 eV in comparison to $\sim$ 8.0 eV for dehydrated DNA, allow us to characterize the DNA 
molecule as a narrow band gap semiconductor. The main factors enhancing the appearance of 
these unbound electrons are linking of the nucleobases with water molecules and intrastrand 
interactions between neighboring nucleobases. The main contribution to the conductivity 
comes from the intrastrand transfer of unbound $\pi$ electrons between purines because of
quite strong overlapping of their orbitals ($S_{C5(\rm{G_1(A_1}))\rightarrow}$$_{C5(\rm{G_2(A_2}))}
\approx\pm$ 0.025) and a small energy gap $\epsilon_{C5(\rm{G_1(A_1)})}-\epsilon_{C5(\rm{G_2(A_2)})}
=0.33$ eV. Therefore, at room temperature sufficient hydration of DNA can lead to an increase 
of the conductivity by up to 10$^3$ times, but at low temperature the conductance properties of the
poly(dG)-poly(dC) and poly(dA)-poly(dT) sequences should be significantly different because 
of the orbital symmetry rule. As an example, because of the out-of-phase interaction of the
wavefunction of the corresponding $\pi$ unbound electrons, the conductivity of the poly(dG)-poly(dC) 
chain is expected to be much larger than that for the poly(dA)-poly(dT) structure, where these
wavefunctions are in-phase. The pairing of the unbound $\pi$ electrons from the nearest-neighbor 
intrastrand bases within the hydrated DNA chain can be with anti-parallel or parallel spins,
that is a source of correspondent diamagnetic and paramagnetic behavior of DNA in a magnetic field.

\section{Acknowledgment}
The work was supported by the Canada Research Chairs
Program and the NSERC Discovery Grant.\\


\begin{thebibliography}{99}
\bibitem{revmod}
R.G. Endres, D.L. Cox, R.R.P. Singh, {\it Rev.
Mod. Phys.} {\bf 2004}, {\it 76}, 195 -- 214.
\bibitem{endres}
M. Xu, R.G. Endres, Y. Arakawa, {\it Small} {\bf 2007}, {\it 3},
1539 -- 1543. 
\bibitem{ratner}
M.A. Ratner, {\it J. Phys. Chem.} {\bf 1990}, {\it 94}, 4877-4883.
\bibitem{lewis122}
F.D. Lewis, R.S. Kalgutkar, {\it J. Am. Chem. Soc.} {\bf 2000}, {\it 122}, 12346-12351.
\bibitem{conw}
E. Conwell, {\it Top. Curr. Chem.} {\bf 2004}, {\it 237}, 73-101.
\bibitem{gieN}
B. Giese, J. Amaudrut, A.K. K\"ohler, M. Spormann, S. Wessely, {\it Nature} {\bf 2001},
{\it 412}, 318-320.
\bibitem{bixon2}
Bixon, M.; Jortner, J. {\it J. Phys. Chem.A} {\bf 2001}, {\it 105}, 10322-10328.
\bibitem{siguyam}
Sugiyama, H.; Saito, I. {\it J. Am. Chem. Soc.} {\bf 1996}, {\it 118}, 7063-7068.
\bibitem{voit}
Voityuk, A.A.; R\"osch, N.; Bixon, M.; Jortner, J. {\it J. Phys. Chem. B} {\bf 2000},
{\it 104}, 9740-9745.
\bibitem{troisi}
Troisi, A.; Orlandi, G. {\it Chem. Phys. Lett.} {\bf 2001}, {\it 344}, 509-518.
\bibitem{lewis}
Lewis, F. D.; Zhu, H., Daublain, P., Fiebig, T.;  Raytchev, M.; Wang, Q.; Shafirovich, V.
{\it J. Am. Chem. Soc.} {\bf 2006}, {\it 128}, 791-800.
\bibitem{barton}
O'Neill, M.A.; Barton, J.K. {\it Proc. Natl. Acad. Sci. USA}  {\bf 2002}, {\it 99}, 16543-16550.
\bibitem{lewisang}
Lewis, F. D.; Zhu, H.; Daublain, P.; Cohen, B.; Wasielewski, M. R. {\it Angew. Chem. Int. Ed.}
{\bf 2006}, {\it 45}, 7982-7985.
\bibitem{book}
Chakraborty, T. (Ed.), {\it Charge Migration in DNA} (Springer, Berlin, 2007).
\bibitem{nanobio}
S. Roy, H. Vedala, A. Datta Roy, D.-H. Kim, M. Doud, K. Mathee, H.-K. Shin, 
N. Shimamoto, V. Prasad, W. Choi, {\it Nano Lett.} {\bf 2008}, {\it 8}, 26 -- 30.
\bibitem{yoo}
Yoo, K.-H.; Ha, D. H.; Lee, J.-O.; Park, J. W.; Kim, J.; Kim, J. J.; Lee, H.-Y.;
Kawai, T.; Choi, H. Y. {\it Phys. Rev. Lett.} {\bf 2001}, {\it 87}, 198102.
\bibitem{kasumov}
Kasumov, A. Y.; Kociak, M.; Gueron, S.; Reulet, B.; Volkov, V. T.;
Klinov, D. V.; Bouchiat, H. {\it Science} {\bf 2001}, {\it 291}, 280-282.
\bibitem{kawai}
Taniguchi, M; Kawai, T. {\it Physica E} {\bf 2006}, {\it 33} 1-12.
\bibitem{porath}
Porath, D.; Bezryadin, A.; deVries, S.; Dekker, C. {\it Nature} {\bf 2000}, 
{\it 403}, 635-638.
\bibitem{jortner}
Jortner, J.; Bixon, M.; Langenbacher, T.; Michel-Beyerle, M. E.
{\it Proc. Natl. Acad. Sci. USA} {\bf 1998}, {\it 95} 12759-12765.
\bibitem{berlin}
Berlin, Y. A.; Burin, A. L.; Ranter, M. A. {\it J. Phys. Chem. A} {\bf 2000},
{\it 104}, 443-445.
\bibitem{kawai1}
Ha, D. H.; Nham, H; Yoo, K.-H.; So, H.-m.; Lee, H.-Y.; Kawai, T.
{\it Chem. Phys. Lett.} {\bf 2002}, {\it 355}, 405-409;
\bibitem{other}
Yamahata, C.; Collard, D.; Takekawa, T.; Kumemura, M.; Hashiguchi, G.; Fujita, H.
{\it Biophys. J.} {\bf 2007}, {\it 94}, 63-70;
Kleine-Ostmann, T.; J\"ordens, C.; Baaske, K.; Weimann, T.; Hrabe de Angelis, M.; 
Koch, M. {\it Appl. Phys. Lett.} {\bf 2006}, {\it 88}, 102102.
\bibitem{tuuk}
S. Tuukkanen, A. Kuzyk, J.J. Toppari, V.P. Hyt\"onen, T. Ihalainen,
{\it Appl. Phys. Lett} {\bf 2005},{\it 87}, 183102.
\bibitem{bouch}
Nakamae S., Cazayous M., Sacuto A., Monod P., Bouchiat H. {\it Phys. Rev. Lett.} 
{\bf 2005}, {\it 94}, 248102.
\bibitem{bouch1}
Nakamae S., Cazayous M., Sacuto A., Monod P., Bouchiat H. {\it Phys. Rev. Lett.} 
{\bf 2006}, {\it 96}, 089802.
\bibitem{berash2}
Berashevich, J.A; Chakraborty, T. {\it Chem. Phys. Lett.} {\bf 2007}, {\it 446}, 
159-164.
\bibitem{siebelas}
Senthilkumar, K.; Grozema, F. C.; Guerra, C. F.; Bilkelhaupt, F. M.; Lewis, D.;
Berlin,Y.A.; Mark A. Ratner, M.A.; Siebbeles, L. D. A.  {\it J. Am. Chem. Soc.} 
{\bf 2005}, {\it 127}, 14894-14903.
\bibitem{macia}
Roche, S.; Macia. E. {\it Mod. Phys. Lett. B.} {\bf 2004},{\it 18}, 847-871.
\bibitem{wang}
Wang, X. F.; Chakraborty, T. {\it Phys. Rev. Lett.} {\bf 2006}, {\it 97}, 106602.
\bibitem{berash}
Berashevich, J.; Bookatz, A. D.; Chakraborty, T. {\it J. Phys.:Cond. Matt.} 
{\bf 2008}, {\it 20}, 035207.
\bibitem{new_my}
Berashevich, J.A; Chakraborty, T. {\it J. Chem. Phys.} {\bf 2008}, {\it 128}, 
235101.
\bibitem{berash1}
Berashevich, J.A; Chakraborty, T. {\it J. Phys. Chem. B} {\bf 2007}, {\it 111}, 
13465-13471.
\bibitem{shuster}
Barnett, R.N.; Cleveland, C.S.L.; Landman, U.; Boone, E.; Kanvah, S.; Schuster, G.B.
{\it J. Phys. Chem. A} {\bf 2003}, {\it 107}, 3525-3537.
\bibitem{parinel}
Gervasio, F.L.; Carloni, P.; Parinello, M.; {\it Phys. Rev. Lett.} {\bf 2002}, 
{\it 89}, 108102.
\bibitem{gunst}
Bonvin, A.M.J.J.; Sunnerhagen, M.; Otting, G.; van Gunsteren W.F.
{\it J. Mol. Biol.} {\bf 1998}, {\it 282}, 859-873.
\bibitem{jaguar}
a) Jaguar, version 6.5. Schr\"odinger. LLC: New York, NY, 2005;
b) Besler, B. H.;  Merz, K. M.; Kollman P. A.{\it J. Comput. Chem.} {\bf 1990}, 
{\it 11}, 431-439.
\bibitem{becke}
Becke, A.D. {\it J Chem. Phys.}  {\bf 1993}, {\it 98}, 5648-5652.
\bibitem{nbo}
http://www.chem.wisc.edu/$\sim$nbo5
\bibitem{sun}
Sun, G.; Nicklaus, M.C. {\it Theor. Chem. Acc.} {\bf 2007}, {\it 117}, 323-332.
\bibitem{book1}
W. Saenger, Principles of nucleic acid structure. New York: Springer-Verlag, 1984. 
p. 556.
\bibitem{rauk}
Rauk A. Orbital interaction theory of organic chemistry. Wiley-interscience. 
2001 2nd ed.
\bibitem{fiebig}
Buchvarov, I.; Wang, Q.; Raytchev, M.; Trifonov, A.; Feibig, T. {\it Proc. 
Natl. Acad. Sci. USA} {\bf 2007}, {\it 104}, 4794-4797.
\bibitem{frohne}
Rist, M.; Wagenknecht, H.-A.; Feibig, T. {\it ChemPhysChem} {\bf 2002}, {\it 104}, 
704-707.
\bibitem{spintro}
Wang, X.F.; Chakraborty, T. {\it Phys. Rev. B} {\bf 2006}, {\it 74}, 193103; 
Zwolak, M.; Di Ventra, M. {\it Appl. Phys. Lett. 81} {\bf 2002}, {\it 81}, 925-927.
925 , X.F.;2002
\end{thebibliography}
\end{document}